\documentclass[12 pt]{article}
\usepackage{amssymb,amsmath}
\usepackage{enumerate}
\usepackage{graphicx}
\usepackage{subcaption}
\usepackage{wrapfig}
\usepackage[letterpaper, margin=1in]{geometry}
\usepackage{setspace}
\usepackage{etoolbox}
\usepackage{authblk}
\usepackage{url}
\usepackage{caption}
\usepackage{bm} 
\preto\table{\par\singlespacing}

\usepackage{array} 
\newcolumntype{C}[1]{>{\centering\let\newline\\
  \arraybackslash\hspace{0pt}}m{#1}}
  
\begin{document}

\title{Cell Line Classification Using Electric Cell-substrate \\Impedance Sensing (ECIS)}
\author[1]{Megan L. Gelsinger}
\author[2]{Laura L. Tupper}
\author[3]{David S. Matteson}

\affil[1]{\footnotesize Department of Statistics and Data Science, Cornell University, Ithaca, NY 14853 (email: \texttt{mlg276@cornell.edu})}
\affil[2]{\footnotesize Department of Mathematics and Statistics, Williams College, 33 Stetson Court, Williamstown, MA 01267 (email: \texttt{llt1@williams.edu})}
\affil[3]{\footnotesize Departments of Statistics and Data Science and Social Statistics, Cornell University, Ithaca, NY 14853 (email: \texttt{matteson@cornell.edu})}

\maketitle

\singlespacing
\begin{abstract}
We consider cell line classification using multivariate time series data obtained from electric cell-substrate impedance sensing (ECIS) technology. The ECIS device, which monitors the attachment and spreading of mammalian cells in real time through the collection of electrical impedance data, has historically been used to study one cell line at a time.  However, we show that if applied to data from multiple cell lines,  ECIS can be used to classify unknown or potentially mislabeled cells, which may help to mitigate the current crisis of reproducibility in the biological literature. We assess a range of approaches to this new problem, testing different classification methods and deriving a dictionary of 29 features to characterize ECIS data. Our analysis also makes use of simultaneous multi-frequency ECIS data, where previous studies have focused on only one frequency. In classification tests on fifteen mammalian cell lines, we obtain very high out-of-sample accuracy. These preliminary findings provide a baseline for future large-scale studies in this field.  
\end{abstract}

{\bf Keywords: } biophysics, classification analysis, multivariate time series, supervised learning 

\doublespacing
\section{Introduction}
Bioelectrical impedance cell culture platforms, such as ECIS, offer scientists the valuable opportunity to study a variety of cell characteristics, such as attachment, growth, and death, in a non-invasive, real-time environment~\cite{ramasamy2014drug}.  Although ECIS technology has been in use since the 1990's, a major application of ECIS data has not yet been formally explored: its potential in identifying unknown or mislabeled cells. Better cell line identification can help address the current scientific reproducibility crisis that results, in part, from the use of misidentified cells in biological research. We introduce the foundations for such a tool, leveraging the multivariate data provided by ECIS technology and insights gained through statistical classification analysis.  While the majority of this manuscript is devoted to our analysis, we begin with a brief introduction to ECIS and to previous work using ECIS technology and data.      
 
In the 1990's, ECIS was introduced as a label-free, real-time, electrical impedance-based method for studying cell behavior in tissue culture~\cite{giaever1991micromotion}.  In the introductory paper, researchers provided both the models and a basic blueprint for the device needed to implement ECIS.  The basic structure they proposed remains in use today. It consists of a typical cell culture plate, in which each well is fitted with a gold electrode, allowing for the passage of electrical currents at various AC frequencies.  After application of such currents, the device measures the electrical impedance associated with each current as a function of time and divides it into its components, resistance and capacitance, allowing for live monitoring of cell activity.  

\begin{figure}[h!]
\begin{center}
\includegraphics[width = \textwidth]{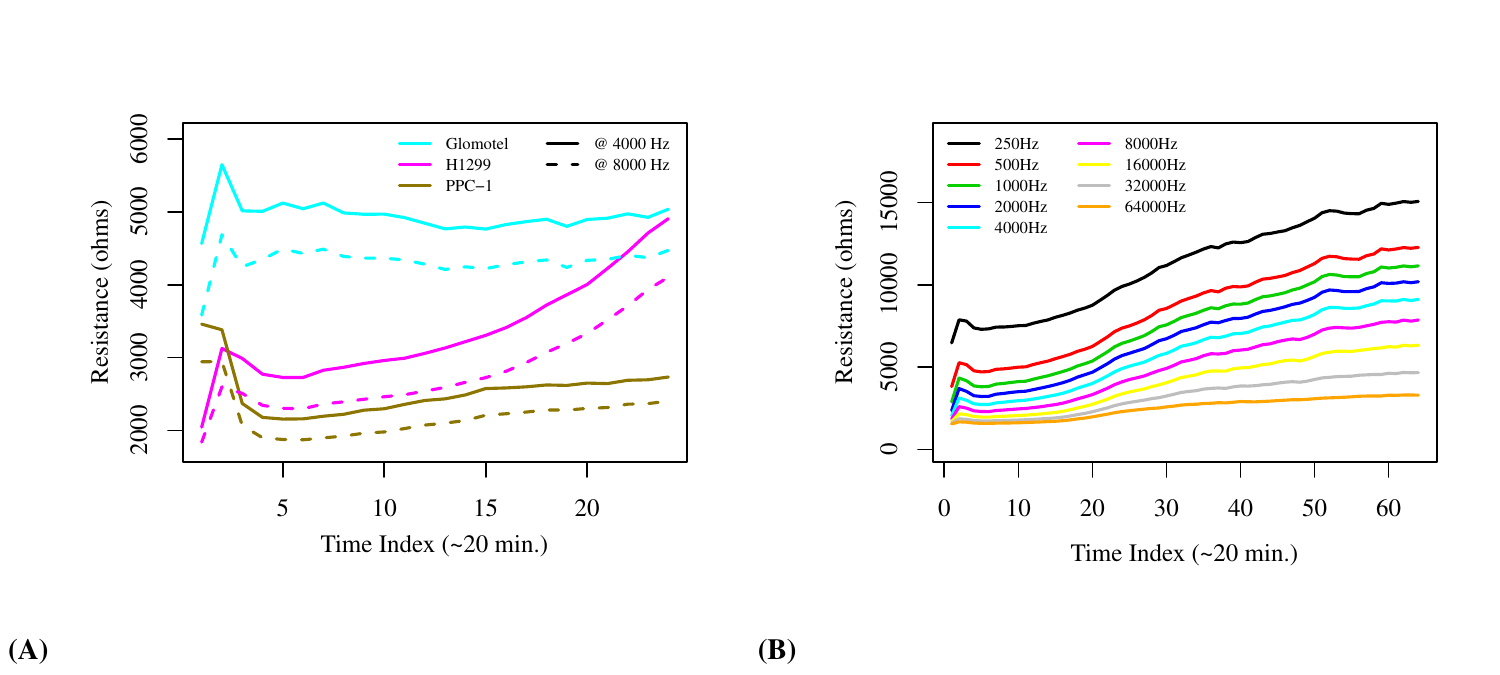}
\end{center}
\caption{{\bfseries Visualizing ECIS Data:} Resistance measurements as recorded by an ECIS\textsuperscript\textregistered device.  (A) A single response for each of three different cell lines, each exposed to two different AC frequencies. (B) Response of the H1299 cell line to all nine AC frequencies.}
\label{threelines}
\end{figure} 

In seeking to expand the scope of ECIS technology to include cell identification and classification, we continue a long history of finding innovative ways to use the originally proposed design~\cite{giaever1991micromotion}.  Since its introduction, researchers in biotechnology have studied ways of improving the overall efficiency and effectiveness of ECIS devices through experimenting with the design scheme~\cite{pradhan2012characterization}. Others in the field have constructed specially tuned versions of an ECIS device for specific applications in cancer cell and clinical diagnostic research~\cite{mishra2005chip, pradhan2014electric}.    

Other studies have used the technology to monitor the activity of cells grown in tissue culture. Early research focused on showing that ECIS measurements could be associated with known biological characteristics of a particular cell line from the attachment and spreading phases of growth \cite{mitra1991electric, lo1993monitoring}. More recently, others have used ECIS to study a cell line before and after an event, such as exposure to cytotoxins, and evaluate induced differences \cite{opp2009use, lovelady2009detecting}.  A more comprehensive overview of the use of ECIS in the biological literature has been detailed elsewhere~\cite{lukic2015impedimetric}.

Previous studies have focused on only one or two cell lines at a time, and their analysis has neglected the full potential of multiple frequency-time measurements available through a standard ECIS device. One group considered two different cell lines, one cancerous and one healthy, but their analysis relied on only one frequency of data and did not extend to additional cell lines \cite{lovelady2007distinguishing}.  Others considered multiple frequencies in their analysis of cytotoxic effects on cell morphology, but they limited their use of the multivariate data to a single time index and only one cell line \cite{opp2009use}.  While these approaches have been sufficient for the scope of studies to date, we consider the full breadth of information available from parallel measurements at multiple frequencies. This additional information allows new types of analysis, including new approaches for cell identification.  

The main application of ECIS data we consider is cell line classification.  Biological studies often rely upon cultured mammalian cells, which are known to mutate or become misidentified during the life of an experiment.  When such anomalies go unnoticed, studies report erroneous results, contributing to the multi-billion dollar irreproducibility problem \cite{freedman2015economics}.  We focus on ECIS, rather than alternative technologies such as short tandem repeat (STR) analysis, due to ECIS' advantages in terms of time and money. ECIS data requires only 24-48 hours to collect and analyze, as opposed to the weeks required by alternative methods that involve shipping out samples for analysis at outside facilities.  In the long run, ECIS devices also prove cost effective: after an up-front purchase of the device (\$3000-\$5000), the user must only invest \$10 per test for a disposable tray.  In contrast, STR costs around \$100 for each test, in addition to the cost of lost time waiting for the test results \cite{keese_gelsinger_2017}.           

While previous studies, such as those cited above, have suggested specific features to help characterize single cell lines using ECIS data, none have employed statistical techniques to quantify the ability of potential characteristic features to classify multiple cell lines at once; the typical statistical analysis in these studies only considered $t$-tests and $F$-tests to measure differences in the mean value of a specific feature across populations.  This is fundamentally different than performing classification based on a feature, as the latter technique simultaneously measures the characteristic across all populations, and quantifies how well it separates and identifies the groups.  Our work determines which features of multi-frequency ECIS data are most effective in a given classification problem; this also suggests features that may be most useful in future analyses.

While we ultimately aim to develop an integrated system to perform cell line classification, that will require much more data collection, and we limit the scope of this manuscript to a preliminary analysis.  We extract many previously considered cell-specific ECIS characteristics from a standard ECIS\textsuperscript\textregistered device, measure them across both time and AC frequency, and quantify their classification performance through established statistical techniques, including classification trees; regularized discriminant analysis (RDA); and two special cases of RDA, linear discriminant analysis (LDA) and quadratic discriminant analysis (QDA).     

The remainder of this paper provides the details of our study.  In Sections 2 and 3, we describe our dataset and feature specification, respectively.  In Section 4, we discuss the classification analysis results, and in Section 5 we offer conclusions, and discuss avenues for future work. 

\section{ECIS Data and Experimental Design}\label{data}
The dataset considered throughout this paper was provided by Applied BioPhysics, the proprietor of ECIS\textsuperscript\textregistered, as recorded by one of their 96-well ECIS\textsuperscript\textregistered Z$\theta$ devices.  It consists of fourteen replicates of fifteen different mammalian cell lines grown on two different serums (gel and bovine serum albumin (BSA)).  The data will be grouped (and compared) by serum type in our forthcoming analysis.  Each plate was exposed to nine different electrical current frequencies over the course of 20 hours, resulting in 64 approximately evenly spaced measurement times.  The fifteen different cell lines are: HUTU80, CHOK1, MDCKII, LCELL, WI-38, VA-13, NRK, PPC-1, DuPro, Glomotel, BSC1, DU145, H1299, NIH3T3, and PC-3.  The nine AC frequencies measured are: 250, 500, 1000, 2000, 4000, 8000, 16000, 32000, and 64000 Hz.  Each 96-well plate held three different cell lines; for each line, we considered data from one plate.  Impedance measurements were recorded approximately once every 20 minutes across all nine AC frequencies (see Figure~\ref{threelines} (B)). The impedance measurements were decomposed into resistance and capacitance according to the originally proposed model~\cite{giaever1991micromotion}.  

Notationally, we can represent the data as follows: $Y_i^{k,s}(t_\ell,\nu_r).$
Here, $Y$ is the resistance value at the given parameter settings.  We enumerate all cell cultures in the dataset, regardless of type, with an overall index $i = 1,...,420$.  Next, $k = 1,..., 15$ represents the index of the cell line in the vector (HUTU80, CHOK1, ..., PC-3), as listed above, and $s = 1, 2$ represents the serum (gel, BSA).  We note that the cell line and serum type are implicitly encoded in the overall index, $i$; likewise, we may suppress $k$ and $s$ without loss of information.  The data depends functionally on $t_\ell$ the discretized time index, with $\ell=1,...,64$ as described above.  It also depends functionally on the discretized input frequency $\nu_r$, with $r=1,...9$, corresponding to the index of the frequency vector $(250, 500,..., 64000),$ as listed above.  This notation helps us understand subsets of the data, which we will reference when describing our classification features.  For example, $\{Y_i^{11, 2}(t_9, \nu_5)\}\equiv Y_{1:14}(t_9, \nu_5)$ corresponds to the resistance of the BSA-treated BSC1 cells at time index nine (hour three) and 4000 Hz.      

\section{Feature Specification}
\subsection{Technical Review}\label{sec:lit}
Many of the characteristics identified in the literature as varying by cell line or cell condition were extracted from data obtained during confluence, or the steady-state portion of growth following the initial growth and spreading stages.  During this phase, the well is completely covered with cells, allowing for only minute movements.  Data collection commonly began after 20-24 hours, once cells had reached confluence, and continued for about 20 hours at a fine sampling frequency, ranging from every two minutes to every second depending on the study.  With this resolution of data, researchers were able to extract features such as: 
\begin{itemize}
\item the slope parameter $\beta$ characterizing a least-squares straight-line fit to a log-log plot of power spectrum versus frequency $\nu$~\cite{lo1993monitoring,opp2009use,lovelady2009detecting,lovelady2007distinguishing,sapper2006cell};   
e.g., Brownian noise displays an $\nu^{-2}$ power law, with $\beta=2$; 
\item the sample kurtosis of the first differenced resistance time series, a measure of distributional shape~\cite{lovelady2007distinguishing};
\item the first $e^{-1}$ crossing of the autocorrelation function of the noise time series from confluence onward (extracted from the resistance time series over the same time period), to estimate its exponential decay rate~\cite{lovelady2009detecting,lovelady2007distinguishing}.
\end{itemize}

\begin{figure}[h!]
\begin{center}
\includegraphics[width = .8\textwidth]{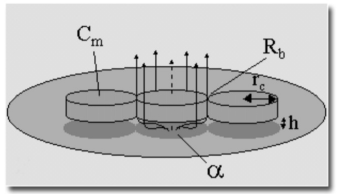}
\end{center}
\caption{{\bfseries Visualizing the ECIS Model:} Image obtained from Applied BioPhysics' website (biophysics.com).  In the ECIS model, cells are viewed as disks, characterized by parameters R$_b, \alpha$, $r_c$,C$_m$, and $h$.}
\label{fig:ecismodel}
\end{figure}

Two other post-confluence features suggested in previous works were R$_b$ and $\alpha$.  R$_b$ reflects the barrier resistance between cells, whereas $\alpha$ reflects the constraint on current flow beneath the cells (see Figure~\ref{fig:ecismodel} for visualization).  R$_b$ and $\alpha$ are defined by the following relationships: 
\begin{align}
\frac{1}{Z_c(\nu)} &= \frac{1}{Z_n(\nu)}\left(\frac{Z_n(\nu)}{Z_n(\nu)+Z_m(\nu)}+\frac{\frac{Z_n(\nu)}{Z_n(\nu)+Z_m(\nu)}}{\frac{\gamma r_c}{2}\frac{I_0(\gamma r_c)}{I_1(\gamma r_c)}+R_b\left(\frac{1}{Z_n(\nu)}+\frac{1}{Z_m(\nu)}\right)}\right);\\
\gamma r_c &= r_c\sqrt{\frac{\rho}{h}\left(\frac{1}{Z_n(\nu)}+\frac{1}{Z_m(\nu)}\right)} = \alpha \sqrt{\frac{1}{Z_n(\nu)}+\frac{1}{Z_m(\nu).}}
\end{align}
For frequency $\nu$, $Z_c(\nu)>0$ is the impedance (per unit area) of the cell-covered electrode, $Z_n(\nu)$ is the impedance of the cell-free (empty, reference) electrode, $Z_m(\nu)$ is the membrane impedance of the cells, $r_c$ is the radius of the cell, $\rho$ is the resistivity of the solution, $h$ is the height of the space between the ventral surface of the cell and the substrate, and $I_0$ and $I_1$ are the modified Bessel functions of the first kind of order 0 and 1~\cite{giaever1991micromotion}. 

\bigskip

An ECIS\textsuperscript\textregistered device is incapable of recording all of the parameters in equations (1) - (2).  Instead, it records only $Z_c(\nu)$ and $Z_n(\nu)$ over a set of frequencies $\{\nu_1,...,\nu_w\}$, with $w\geq3$, and estimates the values of $Z_m(\nu)$, $R_b$ and $\alpha$ which fit the data best. Since these optimal values cannot be computed analytically, the ECIS\textsuperscript\textregistered software uses the Nelder-Mead downhill simplex method to select a series of parameter values.  From there, the program calculates the mean squared error between the measured and modeled frequency scan values, and returns the set of parameter values which minimize this error. The absolute error is normalized to the cell-free reference, so the frequencies are appropriately weighted. It is important to note that underlying these equations and calculations is the assumption that the ECIS model can be understood as a series resistor-capacitor circuit, with cells represented by circular discs hovering above (not directly touching) a gold electrode.  We consider $R_b$ and $\alpha$ as characteristics which may differentiate cell lines, as they were considered in other previous studies~\cite{giaever1991micromotion, opp2009use}. 
 
Several previous studies also evaluated features that were specific to the attachment and spreading phases of growth.  In particular, some researchers found that certain cell lines ``peaked higher" and ``increased more rapidly" during these initial stages of growth~\cite{heijink2010characterisation,park2009electrical}.  Additional remarks are very limited as the majority of prior studies concentrated less on attachment and spreading and more on confluence behavior.

The statistical analysis involved in these studies focused on $t$-tests and $F$-tests to look for significant differences in a feature's mean value across populations of interest.  For example, when one group considered whether a particular cell population inoculated with varying degrees of a cytotoxin had differing mean $R_b$ values for each dosage level, they performed Student's $t$-test for two samples with unequal variance, for pairs of dosages, such as 5 $\mu$M versus 10 $\mu$M of cytotoxin~\cite{opp2009use}.  Another example came from the comparison of cancerous and noncancerous cells, in which an $F$-test was conducted to assess the likelihood of the two cell lines coming from the same distribution, given the means and variances of their power slopes $\beta$~\cite{lovelady2007distinguishing}. 

\subsection{Feature Space Specification}
In our study, we included $R_b$ and $\alpha$ as our confluence phase features.  Computationally only one time index worth of data during the confluence phase is needed to approximate these values, making them feasible for our analysis; we obtained twelve observations per frequency during this phase of growth, following the sampling scheme described previously.  Typically, when using a single time index, the data from all measured frequencies are combined to estimate $R_b$ and $\alpha$. Since we had twelve time indices worth of data, though, we averaged the $R_b$ and $\alpha$ at each of these twelve points to get more stable estimates of $R_b$ and $\alpha$.   

We note that since the design for our study did not emphasize the confluence stage of growth, we did not extract any of the other confluence region features mentioned in Section~\ref{sec:lit}. Studies which did look at these additional features had thousands of observations available for a single frequency during confluence for their analysis, compared to our twelve, leading to our decision to concentrate on $R_b$ and $\alpha$.   

To extend our feature space, we also included proxies for those features from the attachment and spreading portions of cell growth mentioned in Section~\ref{sec:lit} using the $Z_c(\nu)$ data collected from the ECIS\textsuperscript\textregistered device.  These included, at each frequency $\nu_r$: 
\begin{enumerate}[i.]
\item End of run resistance value: $\frac{1}{5}\sum_{\ell=60}^{64}Y_i(t_\ell, \nu_r)$ (EOR);
\item Maximum resistance of moving average smoothed time series value: \\$\max_k \left[\frac{1}{5}\sum_{\ell=k-2}^{k+2}Y_i(t_{\ell}, \nu_r), \textup{ for }k = 3,\hdots,62\right]$ (MR);
\item Resistance at two hours ($t_6$): $\frac{1}{5}\sum_{\ell=4}^{8}Y_i(t_\ell, \nu_r)$ (R2h);
\end{enumerate}  
where smoothing was leveraged for stability purposes.  

The utility of using multi-frequency data in our analysis stems from the physics underpinning the system. We note that for the cell layer, the membrane capacitance and the solution resistance are fixed values and not frequency dependent. The path the current will flow, though, is frequency dependent. Hence, changing the frequency changes the path though which it will flow, and thus the overall impedance of the system $(Z_c(\nu))$.  It is known that at lower frequencies, more current flows under and between cells, making it easier to quantify the resistance between the cells. At higher frequencies, the reactance of the cell membrane is low, with the majority of the current flowing capacitively through the membrane, allowing for an easier exploration of the cell-substrate interaction.  Likewise, $Z_c(\nu)$ at different frequencies captures different characteristics of a cell's morphology, making multi-frequency data potentially more useful for classification than data from just a single frequency. It is also important to keep in mind the presence and effect of the gold electrode that the cells are grown on. In particular, at low frequencies the gold interface resistance becomes very large, potentially contributing a large portion of the overall resistance of the system. For these reasons, we use a range of AC frequencies to create a more accurate picture of the various sources of impedance and resistance in the system, especially those that relate to the cell membrane~\cite{giaever1991micromotion}.

Previous studies involving these features evaluated them at only one frequency, typically 4000 Hz.  One of the main advantages of ECIS\textsuperscript\textregistered Z$\theta$, though, is its ability to record multi-frequency data.  Given the physical justification for using this multi-frequency data, we include relevant features (EOR, MR, R2h) at all nine measurement frequencies given in Section~\ref{data}.  This gives us a total of $(3\times9) + 2 = 29$ characteristics in our feature space which we utilize in our classification analysis below.  In particular, we will investigate whether single, pairs of, and/or trios of these features applied to various classification techniques sufficiently differentiate the cell lines.  We focus on this collection of features as we can assess their performance both quantitatively, through numeric output, and qualitatively through visualizations.

\section{ECIS-based Cell Line Classification Analysis}     
Classification algorithms provide predictive accuracy rates which reflect quantitatively how well certain characteristics differentiate the data.  In previous ECIS-related works, none of these algorithms were used to assess features for general cell line classification; instead, $F$-tests and $t$-tests were relied upon to determine whether the mean value of the feature in question differed by group.  We consider a variety of classifiers in our study to provide this missing quantitative evaluation of potential features for cell line classification.  

Classification analysis was performed using several different supervised learning methods on several different combinations of features.  The first ``grouping" of features considered was similar to that of previous works, where each feature was analyzed on an individual basis as a tool for cell line differentiation.  We also considered all pairs and trios of our 29 features to determine if any provided better separation of the data, and if so, by how much.  All combinations of features were evaluated using both classification trees and RDA (with LDA and QDA as special cases).  

As mentioned in Section~\ref{data}, we examined those cells inoculated in gel and BSA serums separately looking to gain insight on the conditions which promote optimal cell line classification.  The analysis that follows provides an in-depth description and visualization of the classification procedures performed on the gel-based cells.  The same procedures were performed on BSA-based cells; we provide a highlighted summary of these results and comparison to the gel-based cell results at the conclusion of this section.   

\subsection{Classification Trees}
We began our analysis of the gel-based cells with classification trees, as they use the most elementary division of the feature space to classify observations into groups.  Classification trees perform an iterated series of binary splits on the data space to create several regions, then ultimately assign each observation in a given region to the most commonly occurring class of training observations in that region.  To create these splits, we select the predictor, or feature, $X_j$ and cutpoint $b$ such that splitting the feature space into regions $\{X|X_j < b\}$ and $\{X|X_j \geq b\}$ leads to the greatest possible improvement in a particular metric.  We note that based on this definition, the splits in classification trees divide the feature space into a series of rectangles in two-dimensional space.   

One of the most common metrics for assessing the quality of a particular split is cross-entropy or deviance, which is defined for a particular region $m$ by
\[
D_m = -\sum_{k=1}^K
\hat{p}_{mk}\log\hat{p}_{mk}.
\]  
Here we assume the data can be divided into $K$ classes, with $\hat{p}_{mk}$ representing the portion of training observations in the $m$th region that are from the $k$th class.  Cross-entropy, therefore, represents a measure of node purity. A region $m$ is ``pure" when $D_m$ assumes a small value.  This occurs when $\hat{p}_{mk}$ is near zero or one for each $k$, meaning nearly none or nearly all of the training observations in the $m$th region are from the $k$th class, respectively.  So, to construct a binary split, we consider all predictors $X_1,...,X_c$ at all possible values $b$, and select the feature and cutpoint such that the sum of the $D_m$'s associated with the two new regions generated by the split is minimized.  A classification tree is built by iteratively performing these splits on training data until reaching a specified node purity threshold or minimum number of observations per node.  A test set can then be classified using the tree, and a predictive accuracy rate computed~\cite{statlearning}.    

Traditionally, all potential features are passed to the classification tree algorithm, which selects those features and associated splits which lead to optimal node purity.  It is often the case in this traditional approach, though, that more than one, two, or three features appear in the splitting rules for the resultant tree, given there are at least that many features provided during the construction phase.  Given our desire to provide a visual accompaniment to our quantitative results, we opt for a non-traditional approach to constructing our classification trees, only providing them with one, two, or three potential features.  This way we can visualize the resultant splits of the features in 1D, 2D, or 3D space.

In our study, $K=15$, corresponding to the 15 different cell lines in the dataset. We construct all possible trees that use one, two, or three of the features described in Section 3.2, such as EOR@2000Hz ($(1/5)\sum_{\ell=60}^{64}Y_i(t_\ell, \nu_4)$)  and R2H@16000Hz ($(1/5)\sum_{\ell=4}^{8}Y_i(t_\ell, \nu_7)$).  For example, one single-feature tree is based on $X_1$=EOR@2000Hz; each split in the tree is based on an optimal cutpoint corresponding to a particular value of this feature. As there are 29 features, we obtain 29 single-feature trees. Next we build trees using each pair of features: in one such tree, for example, we allow node splits on either $X_1$=EOR@2000Hz or $X_2$=R2H@16000Hz. We construct ${29 \choose 2}=406$ such trees, and similarly ${29 \choose 3}=3654$ trees involving three features.  
  
The training data used to grow the trees is obtained by randomly dividing the entire dataset for each cell line in half, using seven of the fourteen available observations as the training set and the remaining seven as the test set.  We construct the classification trees using the \texttt{tree()} function in R on the training set and assess the out-of-sample classification predictive accuracy rate using \texttt{predict.tree()} on the test set.  We repeat this analysis on twenty different random splits of the data into training and testing sets to lessen the effects of overfitting to any particular sample.  Reported accuracy rates for each feature, feature pair, or trio of features are the average over all twenty of these trials.       

\begin{table}[h!]
\begin{center}
\begin{tabular}{C{2cm}C{6cm}C{2.5cm}C{2.5cm}}
 {\bfseries Feature Space Dimension} & {\bfseries Selected Classification Feature(s)} & {\bfseries Out-of-Sample Classification Accuracy} & {\bfseries Approximate Standard Error} \\ 
  \hline
1 & EOR @ 2000 Hz & 0.717 & 0.009 \\ 
   \hline
2 & R2h @ 32000 Hz \hspace{3cm} MR @ 32000 Hz & 0.952 & 0.005 \\ 
   \hline
3 & R2h @ 32000 Hz \hspace{3cm} MR @ 16000 Hz \hspace{3cm} Rb & 0.968 & 0.004 \\ 
  \end{tabular}
\caption{{\bfseries Classification Tree Analytical Results:} Best feature(s) for cell line classification based on twenty trials of classification tree construction using cells grown on gel serum.  Abbreviations are EOR: End of run resistance; MR: Maximum resistance; R2h: Resistance at two hours.  The out-of-sample classification rate is the average over all twenty values obtained for each feature.  The standard error of this average value is reported in column four. These results show that when classifying cell lines, a 2D feature space is significantly more informative than a 1D feature space.  Using three features as opposed to two results in a less dramatic improvement in the out-of-sample classification rate, but an improvement nonetheless.}
\label{tab:tree}
\end{center}
\end{table}  

Table~\ref{tab:tree} reflects the best feature(s) for classification based on the average accuracy rate. 
We note that the features highlighted in Table~\ref{tab:tree} are the best by a small margin; at least 10\% of all the single, pairs and trios of features were within 5\% of the best respective classification rate.  This demonstrates that we have not only one ultimate best feature, but a set of high-performing features that we could use to differentiate the cell lines; it is possible that one such feature in the set could be more easily obtainable than the others, making it more ideal to use than the designated ``best."  This group of high-performing features is important to consider when interpreting a further implication of Table~\ref{tab:tree} --- the results apply to more than just the features highlighted in Table~\ref{tab:tree}; they apply to the entire class of high performing features which cannot be reproduced for space's sake.  The table reveals a large increase in the out-of-sample classification accuracy when using an informative pair of features as opposed to an informative single feature.  There is a less marked improvement when using a strong trio of features versus a strong pair of features.  Lastly, the table suggests that two or three informative features are sufficient to capture the group dynamics of fifteen different cell lines, as the accuracy rates exceed 90\%.  

\begin{figure}[h!]
\begin{center}
\includegraphics[width = \textwidth]{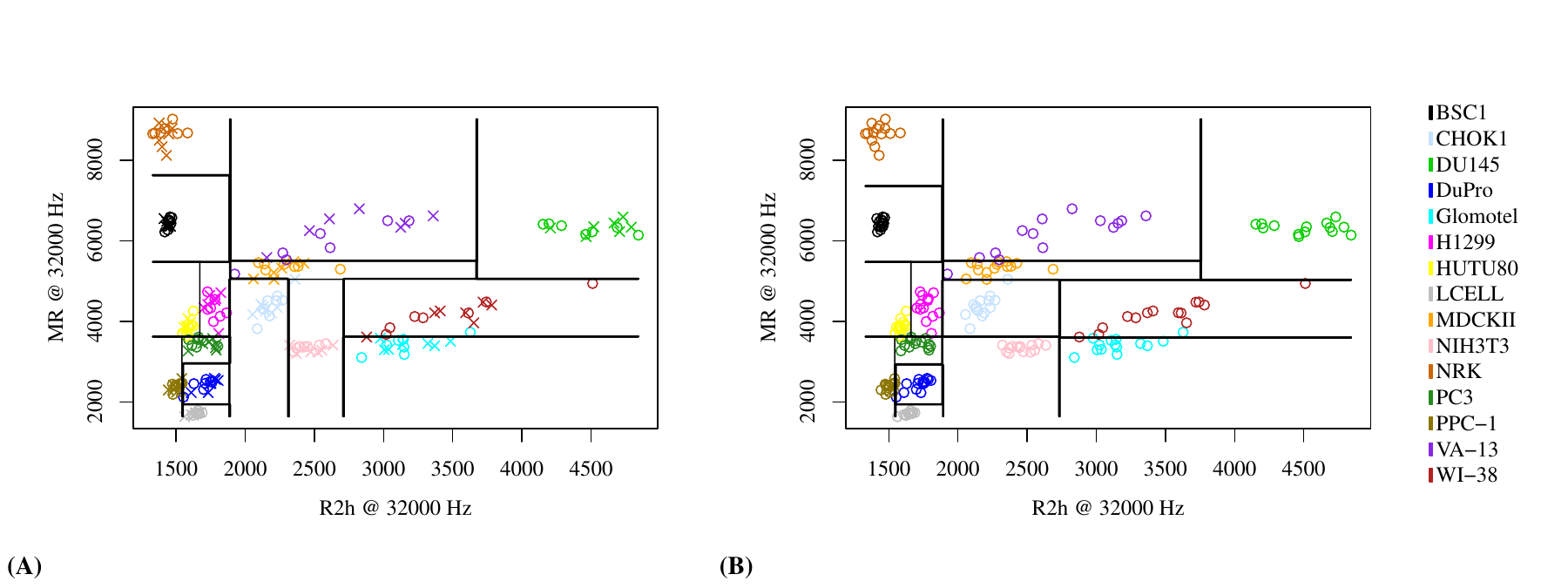}
\end{center}
\caption{{\bfseries Classification Tree Visual Results:} Best pair of features for cell line classification as determined through classification trees using cells grown on gel serum.  (A) Example of training/testing set used to conduct classification analysis.  Training points are represented by hollow circles, test points represented by x's.  Most observations of the same color lie in the same rectangular region, indicating good separation.  This behavior corresponds to the high classification accuracy $(95\%)$ presented in Table~\ref{tab:tree}. (B)  Boundaries created when using all 14 observations in training set (full information).}
\label{fig:tree_best}
\end{figure}

Figure~\ref{fig:tree_best} offers a visualization of the best pair of features proposed in Table~\ref{tab:tree}, and  their ability to separate the data with respect to classification trees.  In panel (A), we see that most of the cells from the same line (represented by points of the same color) reside in the same rectangular region, indicating correct classification.  This behavior reflects the high out-of-sample classification rate (95\%) observed in Table~\ref{tab:tree}. 

\newpage

\subsection{Linear and Quadratic Discriminant Analysis}
While classification trees, and their orthogonal divisions of the feature space, yielded high out-of-sample classification accuracy, we wanted to see if a more flexible separation of the feature space could produce even better results.  Linear and quadratic discriminant analyses are appropriate extensions, as they separate the feature space with linear (not necessarily orthogonal to an axis) and quadratic divisions, respectively.  As both LDA and QDA are special cases of RDA, we reserve the details of their results to Section 4.3 where we discuss RDA at length; this section is meant to provide the mathematical formulation of LDA and QDA for those unfamiliar with the methods.  As a brief note, a normality assumption is associated with each of the following discriminant methods --- that the feature data within each cell line class is normally distributed.  In practice, it has been found that these methods are robust to mild violations of this assumption, i.e., so long as the data is not heavily skewed or multimodal~\cite{odom1991generalization}.  We graphically assessed our data for approximate normality within each class and found no major violations. 

Formally, LDA assumes that observations $X$ of the $k$th class are drawn from the multivariate Gaussian distribution with mean vector $\mu_k$ and covariance matrix $\Sigma:$ ($X\sim \textup{Normal}(\mu_k,\Sigma))$.  Note the lack of subscript on the covariance matrix; LDA assumes that all $K$ classes have the same covariance structure $\Sigma$, a characteristic which distinguishes this method from other classifiers (see QDA below).  The LDA algorithm itself is based on Bayes's theorem.  Assuming a Gaussian density for the $k$th class, the Bayes classifier assigns an observation $X=x$ to the class $\kappa$ for which the posterior probability of $x$ belonging to the $k$th class is maximized.  This corresponds to assigning an observation $X=x$ to the class $\kappa$ for which
\[
\kappa_{\text{Bayes}} = \textup{arg}\max_k\delta_k(x) = \textup{arg}\max_k [x^{T}\Sigma^{-1}\mu_k-\frac{1}{2}\mu_k^{T}\Sigma^{-1}\mu_k  + \log\pi_k],
\]     
where $\pi_k$ is the prior probability of an observation belonging to the $k$th class.  The LDA algorithm estimates $\mu_k$, $\Sigma$ and $\pi_k$ from the training set, which yields estimates for $\hat{\delta}_k(x)$.  Typically, $\hat{\mu}_k$ and $\hat{\Sigma}$ are traditional sample estimates of the mean and covariance, the latter of which might have a scaling constant to mitigate bias.  The prior $\pi_k$ is usually estimated as the proportion of the training data that belongs to the $k$th class.  Likewise, when the sample proportions of each class are the same, as is the case in our study, the $\log\hat{\pi}_k$ terms cancel, leading to
\[\kappa_{\text{LDA}} = \textup{arg}\max_k\hat{\delta}_k(x) = \textup{arg}\max_k [x^{T}\hat{\Sigma}^{-1}\hat{\mu}_k-\frac{1}{2}\hat{\mu}_k^{T}\hat{\Sigma}^{-1}\hat{\mu}_k ].\]
 An observation $X=x^{*}$ from the testing set is then assigned to the $\kappa$th class if $\kappa = \textup{arg}\max_k\hat{\delta}_k(x^{*})$.  The predicted classes using this scheme can then be compared to the true classes to assess classification accuracy~\cite{statlearning}.

QDA is similar to LDA; QDA assumes that observations $X$ of the $k$th class are drawn from the multivariate Gaussian distribution with mean vector $\mu_k$ and covariance matrix $\Sigma_k:$ ($X\sim \textup{Normal}(\mu_k,\Sigma_k))$.  Here too the Bayes classifier assigns an observation $X=x$ to the class $\kappa$ for which the posterior probability of $x$ belonging to the $k$th class is maximized.  This corresponds to assigning an observation $X=x$ to the class $\kappa$ for which
\[
\kappa_{\text{Bayes}} = \textup{arg}\max_k\delta_k(x) = \textup{arg}\max_k \left[-\frac{1}{2}(x-\mu_k)^{T}\Sigma_k^{-1}(x-\mu_k) + \log\pi_k\right].
\]     
Notice here that each of the $k=1,...,K$ classes has its own distinct covariance function, and that the observation $x$ appears in a quadratic form in the Bayes classifier.  These are the defining characteristics of QDA.  The QDA algorithm estimates $\mu_k$, $\Sigma_k$ and $\pi_k$ from the training set similarly to LDA estimates, which yields estimates for $\hat{\delta}_k(x)$,
\[\kappa_{\text{QDA}} = \textup{arg}\max_k\hat{\delta}_k(x) = \textup{arg}\max_k \left[-\frac{1}{2}(x-\hat{\mu}_k)^{T}\hat{\Sigma}_k^{-1}(x-\hat{\mu}_k)\right].\]

  An observation $X=x^{*}$ from the testing set is then assigned to the $\kappa$th class if $\kappa = \textup{arg}\max_k\hat{\delta}_k(x^{*})$.  As above, the predicted classes using the QDA scheme can then be compared to the true classes to assess classification accuracy~\cite{statlearning}.  

\subsection{Regularized Discriminant Analysis}\label{sec:rda}
To see if a more flexible separation of the feature space would yield higher classification rates, we performed classification analysis through the RDA algorithm on all combinations of features for the gel-based cells~\cite{friedman1989regularized}.  This form of discriminant analysis was appealing for two reasons: first, it allows for a ``path" between LDA and QDA which is computationally feasible in high dimensions, and secondly, it has LDA and QDA as special cases.  Not all data conforms exactly to the complete homoscedasticity that LDA demands nor the complete heteroscedasticity that QDA demands.  Instead, their covariance structure decomposes into a mixture of the two extremes, with some shared structure across all $K$ classes and some unique structure within the $k$th class.  RDA analysis allows for such composite covariance structures, including both LDA and QDA as special cases.  The RDA algorithm begins with computing a convex combination of the group-specific QDA) and pooled (LDA) sample covariance matrices over the $k$ groups in the sample via
\[
\hat{\Sigma}_k^{\rho_1} = (1-\rho_1)\hat{\Sigma}_k+\rho_1\hat{\Sigma}
\] 
where $0\leq\rho_1\leq1$ is a regularization parameter.  The second step shrinks this estimator towards a multiple of the identity matrix in the following manner:
\begin{align}
\hat{\Sigma}_{k, \textup{RDA}} = (1-\rho_2)\hat{\Sigma}_k^{\rho_1} + \frac{\rho_2}{p}\textup{tr}(\hat{\Sigma}_k^{\rho_1})I_p,
\end{align}
where $0\leq\rho_2\leq1$ is a second regularization parameter.  Once $\hat{\Sigma}_{k, \textup{RDA}}$ is calculated, performing RDA is analogous to QDA, but with $\hat{\Sigma}_{k, \textup{RDA}}$ substituted for $\hat{\Sigma}_k$. In equation (3), we note that LDA is achieved when $\rho_1 = 1$ and  $\rho_2 = 0,$ which corresponds to creating a series of linear divisions of the feature space.  We also note that QDA is achieved when $\rho_1=\rho_2=0$, which corresponds to creating a series of quadratic divisions of the feature space.  Technically, $0\leq\rho_1,\rho_2\leq2$; however, constructing a model between LDA and QDA, whose divisions of the feature space are somewhere in between linear and quadratic, is only achieved when $0<\rho_1<1$ and $\rho_2 = 0$~\cite{friedman1989regularized}.  Likewise, we restrict our analysis to this setting, exploring a variety of values for $\rho_1$, including the special cases of $\rho_1\in\{0,1\}$, fixing $\rho_2=0$. 

To gain an understanding of the effect of different values of $\rho_1$ on the classification accuracy, we selected values from 0.05 to 0.95 to use in the RDA algorithm on all pairs and trios of features, leaving $\rho_2=0$ fixed.  To run this algorithm in R, we first visually confirmed that our feature data for each cell line was approximately normal.  We next modified the function \texttt{RDA.R} by Aerts and Wilms to accommodate specific values of $\rho_1$ and $\rho_2$~\cite{aertswilms}. The same random sampling scheme used during the classification tree analysis was adopted here, with half of the data constituting the training set, and the other half forming the test set. Here too we repeated our analysis on twenty different random samplings of the data to mitigate any sensitivity to initialization.  Reported accuracy rates for each feature are the average over all twenty of these trials.      

As seen in Supplemental Figure 1, the pair of features with highest out-of-sample classification accuracy is associated with $\rho_1 = 0.1$.  Even so, RDA fit with $0.05\leq\rho_1\leq0.60$ appears to attain a comparable degree of out-of-sample classification accuracy, using the best pair of features associated with each individual $\rho_1$ value.  Thus, it does not appear as though RDA applied to this data set and these pairs of features is extremely sensitive to the $\rho_1$ value, as illustrated in Supplemental Figure 2.  For trios of features, we see similar behavior.  In Supplemental Figure 1, the trio of features with the highest out-of-sample classification accuracy is associated with $\rho_1 = 0.05$; however, RDA seems to perform nearly as well with a wide range of $\rho_1$ values, indicating a lack of sensitivity to this value, apart from it not being zero or one.  It appears as though any form of regularization between LDA and QDA is advantageous to increasing the out-of-sample classification accuracy, with the particular value of $\rho_1$ of lesser importance.   

\begin{table}[b!]
\begin{center}
\begin{tabular}{C{2cm}C{1cm}C{5.8cm}C{2.5cm}C{2.5cm}}
 {\bfseries Feature Space Dimension} & {\bfseries $\bm{\rho_1}$} & {\bfseries Selected Classification Feature(s)} & {\bfseries Out-of-Sample Classification Accuracy} & {\bfseries Approximate Standard Error} \\ 
  \hline
2 & 0.10 & R2h @ 64000 Hz \hspace{3cm} MR @ 16000 Hz & 0.971 & 0.004 \\ 
   \hline
3 & 0.05 & R2h @ 32000 Hz \hspace{3cm} MR @ 4000 Hz \hspace{3cm} MR @ 64000 Hz & 0.993 & 0.002 \\ 
  \end{tabular}
\caption{{\bfseries RDA Analytical Results:} Best features for cell line classification based on twenty trials of RDA with $\rho_1$ as indicated above and $\rho_2=0$.  The cells were grown on gel serum.  The out-of-sample classification rate is the average over all twenty values obtained for each feature.  The standard error of this average value is reported in column five.  The $\rho_1$ values used in the RDA analysis and the associated best classification features correspond to the settings with the highest out-of-sample predictive accuracy in Supplemental Figure 1.  We see an improvement of several percentage points in the classification rates for both pairs and trios of features, as compared to those from the classification trees.}
\label{tab:rda}
\end{center}
\end{table} 

Table~\ref{tab:rda} contains the numerical results associated with the best pair and trio of features identified in Supplemental Figure 1, while Figure~\ref{fig:rda_best} offers a visualization of the feature space characteristic of RDA, somewhere between a linear and quadratic division.  Note that these divisions are quite different from those in Figure~\ref{fig:tree_best}.  In Table~\ref{tab:rda} we see these differences correspond to about a 2 percentage point increase in the out-of-sample classification rate for the best pair of features using RDA versus classification trees.  We also see an increase of about 2 percentage points for the best trio of features using RDA versus classification trees.  We emphasize again, though, that the features highlighted in Table~\ref{tab:rda} are merely members of a larger set of high-performing features.  For example, the pairs (R2h @ 32000 Hz, MR @ 16000 Hz) and (R2h @ 32000 Hz, MR @ 8000 Hz) have classification accuracies only one percentage point worse than the best pair.  So, while not all pairs and trios of features could yield classification rates as high as 97\% and 99\%, respectively, there are a number which could perform within a few percentage points of these values.

\begin{figure}[t!]
\begin{center}
\includegraphics[width = \textwidth]{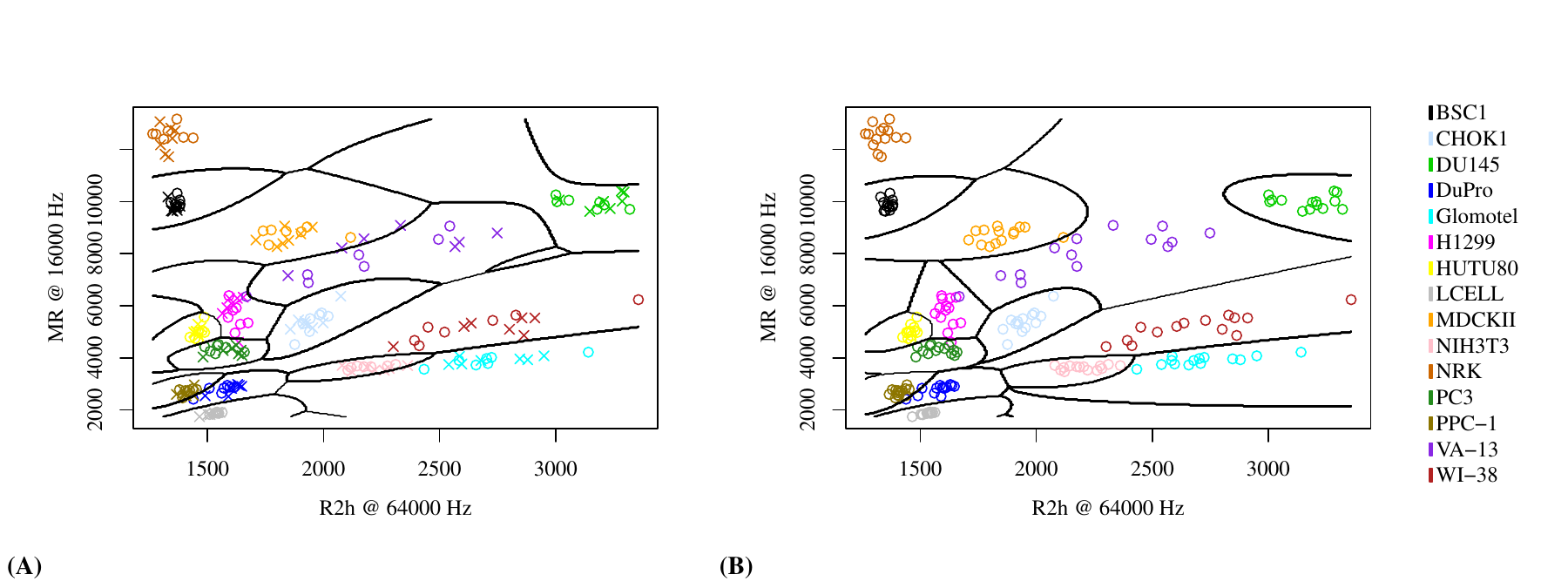}
\end{center}
\caption{{\bfseries RDA Visual Results:} Best pair of features for cell line classification with cells grown on gel serum as determined through RDA with $\rho_1 = 0.10$ and $\rho_2 = 0$ (as indicated in Table 2).  (A) Example of training/testing set used to conduct classification analysis.  Training points are represented by hollow circles, test points represented by x's. This separation corresponds to about 97\% predictive accuracy, as seen in Table~\ref{tab:rda}. (B)  Boundaries created when using all 14 observations in training set (full information).}
\label{fig:rda_best}
\end{figure}

Based on our preliminary analysis on the effect of the $\rho_1$ value on the classification rate, and the low parameter value sensitivity observed, we suspected that the rates associated with LDA and QDA would be similar to those when we specified $\rho_1\in\{0.05,0.1\}$, the optimal parameter values for RDA on trios and pairs of features, respectively.  As LDA and QDA are common classification techniques, with desirable visualization properties, we extended our RDA analysis to account for these two special cases.  The same random sampling and averaging techniques were used here as before.  Instead of using the $\texttt{RDA.R}$ code, we used the \texttt{lda()}, \texttt{predict.lda()}, \texttt{qda()} and \texttt{predict.qda()} functions to allow for evaluation of single features, as well as pairs and trios.  We verified that the results using the modified $\texttt{RDA.R}$ code had the same LDA and QDA results for pairs and trios of features as the \texttt{lda()} and \texttt{qda()} functions.  The numerical and visual results can be found in Supplemental Tables 1-2 and Supplemental Figures 3-4, with highlights summarized below.

Compared to classification trees, we see that LDA and QDA offer nearly uniformly higher out-of-sample classification rates for each combination of features.  RDA, however, outperformed both LDA and QDA.  High-performing features showed some consistency: the best single feature and feature pair under LDA also appeared in the top 10 selected by RDA.  The best single feature and feature trio under QDA also appeared in the top 10 selected by RDA.           

The supplemental figures show how LDA and QDA separate the feature space into linear and quadratic regions, respectively.  A visual comparison of the LDA, QDA, and RDA regions (in Supplemental Figures 3-4 and manuscript Figure 4) shows some similarities; but the classification rates in the corresponding tables in the manuscript and supplemental materials indicate that RDA is still the superior classification method.  Hence, a more flexible classification algorithm, which allows for a division of the feature space somewhere between linear and quadratic, seems to yield the best results for this dataset.   

\subsection{Serum Effect on Classification Rates: BSA vs. Gel}  
The same analysis that was applied to the cells inoculated in gel serum, as described in Sections 4.1 - 4.3, was repeated on those cells inoculated in BSA serum.  Supplemental Table 3 provides the numeric results of the analysis, while Supplemental Figures 5-6 provide visualizations.  The ordering of best to worst classification methods was slightly different for the BSA-based cells, with LDA performing better than QDA, but ultimately, RDA with the optimal $\rho_1$ performing the best overall.     

The BSA-based cells had lower classification rates than the gel-based cells for all classification methods.  Supplemental Figure 6 indicates why BSA-based cells were harder to classify; the plot shows that the observations overlapped more between groups, and, within a particular group, the observations were generally more spread out.  This behavior may also explain why the value of $\rho_1$ in the RDA algorithm had a greater effect on the out-of-sample classification rate for the BSA-based cells (see Supplemental Figure 5); changing the boundaries in Supplemental Figure 6, as dictated by $\rho_1$, would yield more differences than in Supplemental Figure 2.  Supplemental Table 3 also suggests that lower frequencies of data, such as 2000 Hz and 4000 Hz, are more informative for BSA-based cells, as opposed to the higher frequencies, such as 32000 Hz and 64000 Hz, which were selected as the best features for the gel-based cells.  Although the sample size in this study was too small to draw definitive conclusions, these results suggest a slight advantage to using cells grown on gel serum, as opposed to BSA serum.  Biologically, we know that cells bind to surfaces using transmembrane proteins called integrins. Whether a cell binds differently to gelatin or BSA depends upon a variety of factors including the cell's available integrins and whether serum is a component of the medium (which it was in this study).  Given this biological behavior and the initial results in this manuscript, it may be worthwhile to continue to evaluate the differences between these substrates in future experiments.

\subsection{Alternative Approaches}

We considered several additional analyses on the gel-based data to assess the sufficiency of the classification methods and features used in Sections 4.1 - 4.3. To test the performance of the EOR feature versus its component features, the resistance values at time indices (TI) 60 - 64, we created a new feature set replacing EOR with five separate features TI60 through TI64; the remaining original features R2h, MR, Rb, and Alpha were retained.  Conducting classification tree, LDA, and QDA analyses on this new feature set, we found the substitution of TI60 - TI64 for EOR inferior in almost every case (see Supplemental Table 4).  We also performed an analysis using just TI60 - TI64 as features, but this performed uniformly worse (see Supplemental Table 5).  Given this reduced performance, and the fact that the resistance values at the individual time indices are less stable and reliable than the aggregate feature EOR, we further advocate for the use of the latter over the former.     

While trained on a subset of data, classification trees, LDA, QDA, and RDA algorithms are all attractive to use in our application since they all return a set of classification rules that can act on new data without the retention of the training data. Likewise, once generated, it is feasible to share the features with others interested in differentiating their own cell lines into types. This is less straightforward using another popular classification technique, the k-Nearest Neighbor (KNN) method.  Nevertheless, we ran a cross-validated KNN algorithm on our data, following a procedure similar to the one detailed above for the other classification methods.  Results and more details of our procedure are detailed in Supplemental Table 6. Overall, we found KNN underperformed those methods detailed in Sections 4.1 - 4.3.

\section{Conclusion}
These results have established that previous studies have overlooked valuable information by only relying on ECIS data obtained from one frequency, typically 4000 Hz.  While the tables in this manuscript and the supplemental materials provide only examples of the best feature(s), it was clear that neither 4000 Hz, nor any single frequency for that matter, was the sole frequency appearing in the high performing feature sets.  Instead, we found that the most effective trees used feature sets involving multiple frequencies.  We also found that the ``one-at-a-time" feature analysis used in previous work is not sufficient to solve the multiple classification problem.  While only incremental improvement in accuracy was gained from considering trios of features, as opposed to pairs of features, there was a marked improvement when considering pairs versus individual features in this study. As for the best classification method, our results suggest that RDA with a value of $\rho_1$ away from the boundary yields the most accurate results.  We also note that LDA and QDA offer only small improvements over basic classification trees, a margin which decreases as the number of features increases.

Overall, results are encouraging for future work.  We saw clear separation of the various cell lines using our classification features, which previous work had not achieved.  It is possible that the differences in resistance signatures are smaller between a healthy and contaminated version of a cell line, which was the focus of previous work, versus two distinct cell lines, making ours a simpler data set to work with. It is also possible, though, that our features were superior to those used in previous studies since ours were not limited to the confluence region, nor were they restricted to one AC frequency. In general, we found the resistance curves more distinct during the growth phase than the confluence phase, which our features were able to capture and use to better inform the classification procedure.

Since we have demonstrated the value of multi-frequency ECIS data for cell line classification, we plan to extend the experimental design to allow for a broader feature space.  In particular, we will continue to look at cells grown on different serums, and also examine cells wounded by a high-frequency current after reaching confluence~\cite{lukic2015impedimetric}.  We will also obtain more data at a finer temporal resolution during confluence to study the post-confluence features proposed in previous studies that we were unable to address in this work.  As in this study, we will evaluate each feature at multiple frequencies.  We note, though, that in order to apply the discriminant methods used here, any future data must abide by the (approximate) normality condition and have sufficient sample size relative to the feature dimension.  We will also consider the use of functional data features as extensions of this current work, recognizing that examining the time series in their entirety (as curves) may be informative.  Adding these features may enrich the classification algorithm and provide even better cell line separation.  Finally, since our results indicate that two or three features are sufficient for good classification accuracy, a two- or three-dimensional visualization tool could be developed to allow users to view the feature space and classification regions for their specific analysis.  

\section*{Acknowledgements}
The authors would like to thank Applied BioPhysics for providing the data and the technical/biological background for this work.  Financial support from the Cornell University Institute of Biotechnology, the New York State Division of Science, Technology and Innovation (NYSTAR), a Xerox PARC Faculty Research Award, NSF Award DMS-1455172, and Cornell University Atkinson's Center for a Sustainable Future AVF-2017 is gratefully acknowledged as well.  

\section*{Code}
\noindent{All code used to generate the analysis, figures and tables in this manuscript and the Supplementary Materials is publicly available on GitHub:\\ \texttt{https://github.com/megangelsinger/IJBS\_ECIS\_2019.git}}

\bibliographystyle{unsrt}
\bibliography{feature_bib}

\newpage

\captionsetup[figure]{labelfont={bf},name={Supplemental Figure},labelsep=period}
\captionsetup[table]{labelfont={bf},name={Supplemental Table},labelsep=period}
\setcounter{figure}{0} 
\setcounter{table}{0} 

\section*{Supplemental Tables and Figures} 

\begin{table}[h!]
\begin{center}
\begin{tabular}{C{2cm}C{6cm}C{2.5cm}C{2.5cm}}
 {\bfseries Feature Space Dimension} & {\bfseries Selected Classification Feature(s)} & {\bfseries Out-of-Sample Classification Accuracy} & {\bfseries Approximate Standard Error} \\ 
  \hline
1 & EOR @ 1000 Hz & 0.762 & 0.008 \\ 
   \hline
2 & R2h @ 4000 Hz \hspace{3cm} MR @ 8000 Hz & 0.945 & 0.003 \\ 
   \hline
3 & R2h @ 64000 Hz \hspace{3cm} MR @ 2000 Hz \hspace{3cm} MR @ 64000 Hz & 0.974 & 0.003 \\ 
  \end{tabular}
\caption{{\bfseries LDA Analytical Results:} Best feature(s) for cell line classification based on twenty trials of LDA with cells grown on gel serum.  The out-of-sample classification rate is the average over all twenty values obtained for each feature.  The standard error of this average value is reported in column four.  The predictive accuracy rates increased for nearly every combination of features, as compared to the classification tree analysis.  All rates were worse compared to RDA on gel-based cells, when applicable (Table 2 of the manuscript).}
\label{tab:lda}
\end{center}
\end{table} 

\newpage

\begin{table}[h!]
\begin{center}
\begin{tabular}{C{2cm}C{6cm}C{2.5cm}C{2.5cm}}
 {\bfseries Feature Space Dimension} & {\bfseries Selected Classification Feature(s)} & {\bfseries Out-of-Sample Classification Accuracy} & {\bfseries Approximate Standard Error} \\ 
  \hline
1 & EOR @ 1000 Hz & 0.754 & 0.008 \\ 
   \hline
2 & R2h @ 8000 Hz \hspace{3cm} MR @ 32000 Hz & 0.957 & 0.004 \\ 
   \hline
3 & R2h @ 4000 Hz \hspace{3cm} R2h @ 8000 Hz \hspace{3cm} EOR @ 64000 Hz & 0.984 & 0.004 \\ 
  \end{tabular}
\caption{{\bfseries QDA Analytical Results:} Best feature(s) for cell line classification based on twenty trials of QDA with cells grown on gel serum. The out-of-sample classification rate is the average over all twenty values obtained for each feature.  The standard error of this average value is reported in column four.  These results are similar to those from LDA.}
\label{tab:qda}
\end{center}
\end{table}

\newpage

\begin{table}[h!]
\begin{center}
\begin{tabular}{C{2cm}C{2cm}C{5.8cm}C{2.5cm}C{2.5cm}}
 {\bfseries Method} & {\bfseries Feature Space Dimension} & {\bfseries Selected Classification Feature(s)} & {\bfseries Out-of-Sample Classification Accuracy} & {\bfseries Approximate Standard Error} \\ 
  \hline
Tree & 1 & MR @ 1000 Hz & 0.671 & 0.007 \\ 
   \hline
 & 2 & R2h @ 4000 Hz \hspace{3cm} EOR @ 8000 Hz & 0.908 & 0.006 \\ 
   \hline
 & 3 & R2h @ 500 Hz \hspace{3cm} R2h @ 16000 Hz \hspace{3cm} EOR @ 8000 Hz & 0.937 & 0.004 \\ 
   \hline
RDA $(\rho_1= 0.10 )$ & 2 & R2h @ 4000 Hz \hspace{3cm} MR @ 16000 Hz & 0.945 & 0.003 \\ 
   \hline
$(\rho_1= 0.05 )$ & 3 & R2h @ 250 Hz \hspace{3cm} R2h @ 2000 Hz \hspace{3cm} MR @ 16000 Hz & 0.976 & 0.004 \\ 
   \hline
LDA & 1 & EOR @ 2000 Hz & 0.726 & 0.006 \\ 
   \hline
 & 2 & R2h @ 2000 Hz \hspace{3cm} MR @ 2000 Hz & 0.917 & 0.005 \\ 
   \hline
 & 3 & R2h @ 2000 Hz \hspace{3cm} EOR @ 4000 Hz \hspace{3cm} EOR @ 32000 Hz & 0.953 & 0.004 \\ 
   \hline
QDA & 1 & EOR @ 2000 Hz & 0.721 & 0.007 \\ 
   \hline
 & 2 & EOR @ 8000 Hz \hspace{3cm} EOR @ 16000 Hz & 0.941 & 0.005 \\ 
   \hline
 & 3 & R2h @ 4000 Hz \hspace{3cm} R2h @ 32000 Hz \hspace{3cm} Alpha & 0.950 & 0.004 \\ 
  \end{tabular}
\caption{{\bfseries All Analytical Results (BSA-Based Cells):} Best feature(s) for cell line classification based on twenty trials of various classification methods with cells grown on BSA serum. The out-of-sample classification rate is the average over all twenty values obtained for each feature.  The standard error of this average value is reported in column five.  The $\rho_1$ values used in the RDA analysis and the associated best classification features correspond to the settings with the highest out-of-sample predictive accuracy in Supplemental Figure 5. Compared to the results from cells grown on gel serum, we see uniformly lower out-of-sample classification rates and the appearance of Alpha as a best feature.}
\label{tab:bsa}
\end{center}
\end{table}

\newpage

\begin{table}[h!]
\begin{center}
\begin{tabular}{C{2cm}C{2cm}C{5.8cm}C{2.5cm}C{2.5cm}}
 {\bfseries Method} & {\bfseries Feature Space Dimension} & {\bfseries Selected Classification Feature(s)} & {\bfseries Out-of-Sample Classification Accuracy} & {\bfseries Approximate Standard Error} \\ 
  \hline
Tree & 1 & TI60 @ 2000 Hz & 0.710 & 0.007 \\ 
   \hline
 & 2 & R2h @ 32000 Hz \hspace{3cm} MR @ 32000 Hz & 0.952 & 0.005 \\ 
   \hline
 & 3 & R2h @ 32000 Hz \hspace{3cm} MR @ 16000 Hz \hspace{3cm} TI64 @ 4000 Hz & 0.968 & 0.004 \\ 
   \hline
LDA & 1 & TI60 @ 2000 Hz & 0.763 & 0.008 \\ 
   \hline
 & 2 & R2h @ 4000 Hz \hspace{3cm} MR @ 8000 Hz & 0.945 & 0.003 \\ 
   \hline
 & 3 & R2h @ 1000 Hz \hspace{3cm} TI63 @ 2000 Hz \hspace{3cm} TI64 @ 64000 Hz & 0.975 & 0.003 \\ 
   \hline
QDA & 1 & TI60 @ 1000 Hz & 0.767 & 0.008 \\ 
   \hline
 & 2 & R2h @ 8000 Hz \hspace{3cm} MR @ 32000 Hz & 0.957 & 0.004 \\ 
   \hline
 & 3 & R2h @ 4000 Hz \hspace{3cm} R2h @ 8000 Hz \hspace{3cm} TI60 @ 64000 Hz & 0.985 & 0.004 \\ 
  \end{tabular}
\caption{{\bfseries All Analytical Results Using Individual Time Indices Instead of EOR (V.1):} Best feature(s) for cell line classification based on twenty trials of various classification methods with cells grown on gel serum. The possible features in this case were R2h, MR, Rb, Alpha, and TI60 through TI64, where ``TIx" represents the resistance values at time index ``x."  This compares to our previous analysis where we used EOR, the average resistance value over TI60 - TI64, as a possible feature.  The out-of-sample classification rate is the average over all twenty values obtained for each feature.  The standard error of this average value is reported in column five.  Compared to the results using R2h, MR, Rb, Alpha, and EOR (see Table 1 of the manuscript and Supplemental Tables 1 and 2), we find the substitution of TI60 - TI64 for EOR not superior in almost every case.  Since using the resistance values at the individual time indices did not greatly improve the out-of-sample predictive accuracy, the manuscript focuses on our original analysis, as the individual time index features are also less stable.}
\label{tab:tiv1}
\end{center}
\end{table}

\newpage

\begin{table}[h!]
\begin{center}
\begin{tabular}{C{2cm}C{2cm}C{5.8cm}C{2.5cm}C{2.5cm}}
 {\bfseries Method} & {\bfseries Feature Space Dimension} & {\bfseries Selected Classification Feature(s)} & {\bfseries Out-of-Sample Classification Accuracy} & {\bfseries Approximate Standard Error} \\ 
  \hline
Tree & 1 & TI60 @ 2000 Hz & 0.710 & 0.007 \\ 
   \hline
 & 2 & TI60 @ 64000 Hz \hspace{3cm} TI61 @ 4000 Hz & 0.879 & 0.009 \\ 
   \hline
 & 3 & TI61 @ 64000 Hz \hspace{3cm} TI62 @ 4000 Hz \hspace{3cm} TI64 @ 32000 Hz & 0.879 & 0.008 \\ 
   \hline
LDA & 1 & TI60 @ 2000 Hz & 0.763 & 0.008 \\ 
   \hline
 & 2 & TI61 @ 32000 Hz \hspace{3cm} TI62 @ 8000 Hz & 0.895 & 0.005 \\ 
   \hline
 & 3 & TI60 @ 2000 Hz \hspace{3cm} TI60 @ 4000 Hz \hspace{3cm} TI60 @ 16000 Hz & 0.921 & 0.006 \\ 
   \hline
QDA & 1 & TI60 @ 1000 Hz & 0.767 & 0.008 \\ 
   \hline
 & 2 & TI60 @ 4000 Hz \hspace{3cm} TI60 @ 64000 Hz & 0.948 & 0.005 \\ 
   \hline
 & 3 & TI60 @ 8000 Hz \hspace{3cm} TI60 @ 16000 Hz \hspace{3cm} TI60 @ 32000 Hz & 0.945 & 0.007 \\ 
  \end{tabular}
\caption{{\bfseries All Analytical Results Using Only Individual Time Indices Instead of EOR (V.2):} Best feature(s) for cell line classification based on twenty trials of various classification methods with cells grown on gel serum. The possible features in this case were TI60 - TI64, where ``TIx" represents the resistance values at time index ``x."  The out-of-sample classification rate is the average over all twenty values obtained for each feature.  The standard error of this average value is reported in column five.  Compared to the results using R2h, MR, Rb, Alpha, and either EOR or TI60 - TI64 (see Table 1 of the manuscript and Supplemental Tables 1, 2, and 4), we find TI60 - TI64 inadequate for cell line classification.  Based on this and Supplemental Table 4, we find our ``hand-crafted" features more informative for cell line classification than ``raw" features of the data.}
\label{tab:tiv2}
\end{center}
\end{table}

\newpage

\begin{table}[h!]
\begin{center}
\begin{tabular}{C{2cm}C{6cm}C{2.5cm}C{2.5cm}}
 {\bfseries Feature Space Dimension} & {\bfseries Selected Classification Feature(s)} & {\bfseries Out-of-Sample Classification Accuracy} & {\bfseries Approximate Standard Error} \\ 
  \hline
1 & EOR @ 2000 Hz & 0.719 & 0.009 \\ 
   \hline
2 & R2h @ 16000 Hz \hspace{3cm} EOR @ 32000 Hz & 0.939 & 0.005 \\ 
   \hline
3 & R2h @ 2000 Hz \hspace{3cm} R2h @ 16000 Hz \hspace{3cm} MR @ 64000 Hz & 0.981 & 0.003 \\ 
  \end{tabular}
\caption{{\bfseries KNN-CV Analytical Results:} Best feature(s) for cell line classification based on twenty trials of KNN-CV with cells grown on gel serum.  Our implementation of this method mimicked that of the classification tree, LDA, and QDA methods, except for a tuning step prior to fitting the KNN. In this tuning step, we considered 10-fold cross-validation of KNN using values of $k$ from 1 to 20.  The out-of-sample classification rate is the average over all twenty trial (subsample) values obtained for each feature.  The standard error of this average value is reported in column four.  KNN-CV does not perform markedly better than classification trees, LDA, QDA, or RDA.  Since LDA, QDA, and RDA perform nearly as well as or better than KNN-CV and their features can easily be extracted and used without retaining the training data, we advocate focusing on these features in this and future work.} 
\label{tab:knn}
\end{center}
\end{table}

\newpage

\begin{figure}[h!]
\begin{center}
\includegraphics[width = \textwidth]{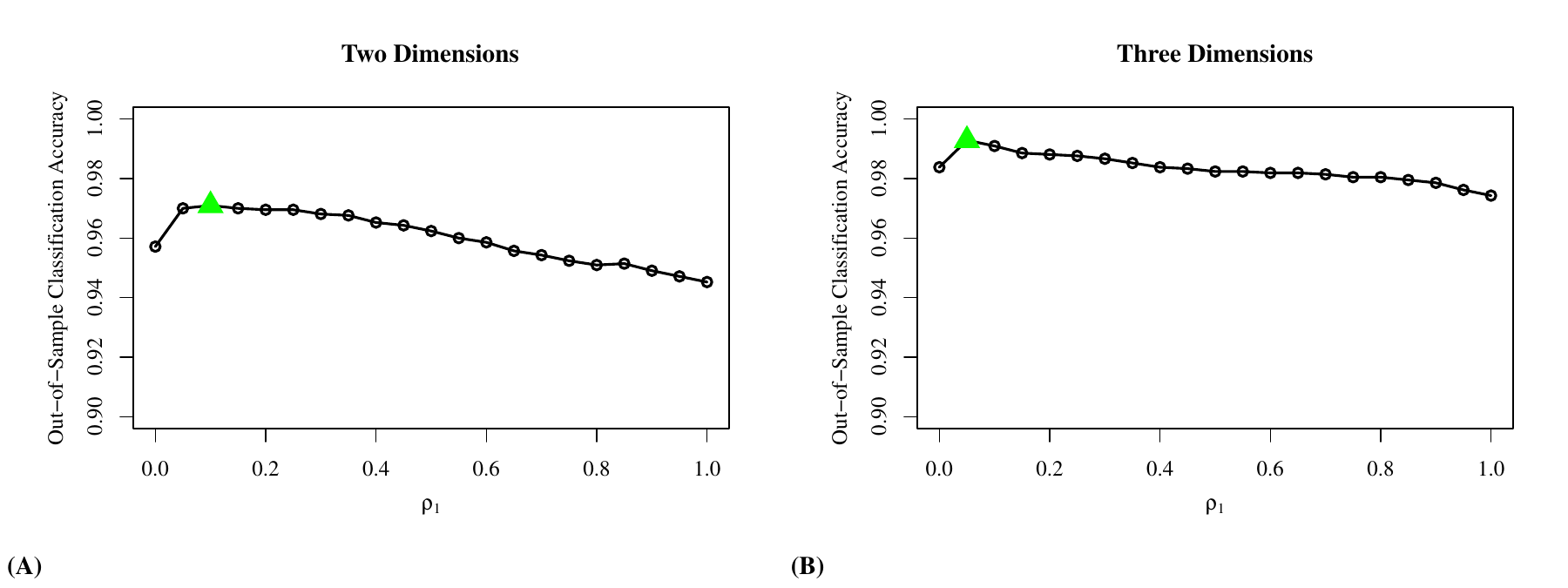}
\end{center}
\caption{{\bfseries Exploration of $\boldsymbol{\rho}_1$ for Gel-Based Cells:} Highest RDA out-of-sample classification accuracy for cells grown on gel serum achieved among all  (A) pairs of features and (B) trio of features for each fixed value of $\rho_1 \in (0, 0.05, 0.1, ..., 0.95, 1)$, holding $\rho_2 = 0$. The maximum for each value of $\rho_1$ is associated with the best corresponding pair (trio) of features.  The maximum out-of-sample classification accuracy across all of the $\rho_1$ values is indicated with a green triangle in the figure.  Note that $\rho_1 = 0$ corresponds to QDA and $\rho_1 = 1$ corresponds to LDA.  We note in (A) and (B) the high maximum out-of-sample classification accuracy achieved for each fixed value of $\rho_1$.  This figure suggests that RDA using our data is not very sensitive to the $\rho_1$ value.} 
\label{rho1s_gel}
\end{figure}

\newpage

\begin{figure}[h!]
\begin{center}
\includegraphics[width = \textwidth]{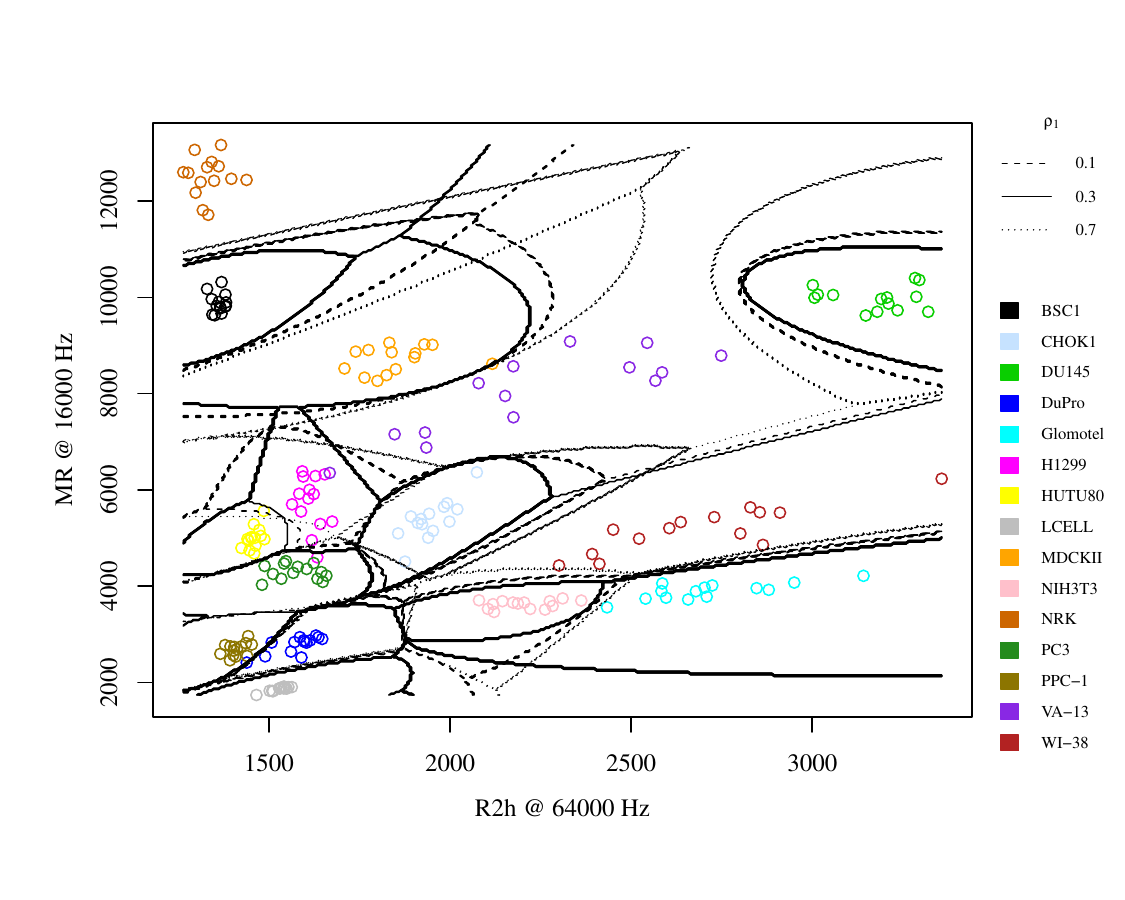}
\end{center}
\caption{{\bfseries RDA Boundary Changes with $\boldsymbol{\rho}_1$:} Comparison of feature space division given three different values of $\rho_1$ in RDA algorithm on the best pair of features indicated in Table 2 of the manuscript, with cells grown on gel serum and $\rho_2=0$ fixed.  Recall that $\rho_1 = 0.1$ is the theoretical best value associated with these features. Even so, we see only slight changes in the boundaries, suggesting the value of $\rho_1$ has little impact on the classification rate of the RDA algorithm in this scheme.  This observation is also echoed in Supplemental Figure 1.}
\end{figure}

\newpage

\begin{figure}[h!]
\begin{center}
\includegraphics[scale = 0.6]{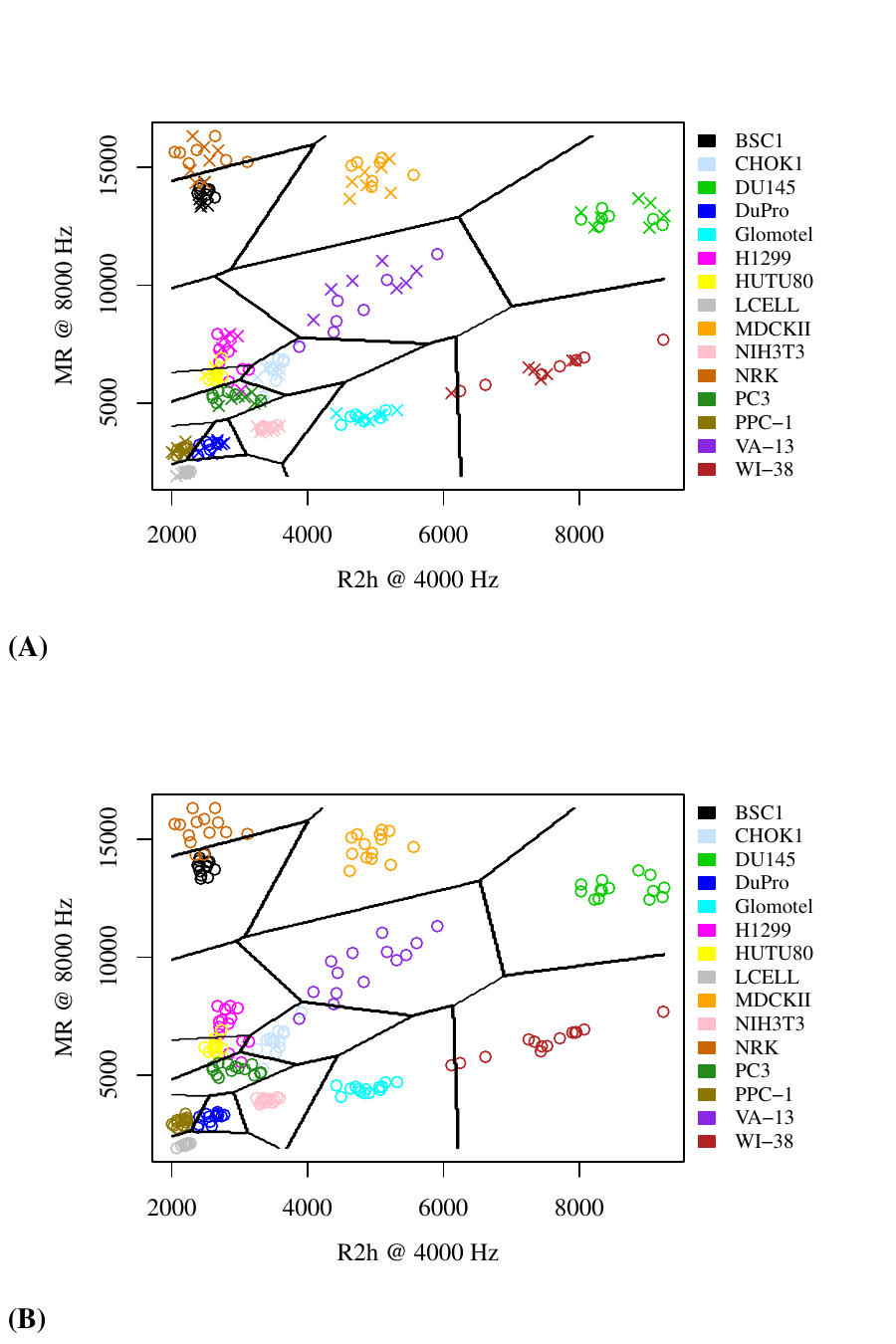}
\end{center}
\caption{{\bfseries LDA Visual Results:} Best pair of features for cell line classification as determined through LDA with cells grown on gel serum.  (A) Example of training/testing set used to conduct classification analysis.  Training points are represented by hollow circles, test points represented by x's.  Most observations of the same color lie in the same oblong region, indicating good separation.  This behavior corresponds to the high classification accuracy $(\approx 95\%)$ presented in Supplemental Table~\ref{tab:lda}.  (B) Boundaries created when using all 14 observations in training set (full information).}
\label{fig:lda_best}
\end{figure}

\newpage

\begin{figure}[h!]
\begin{center}
\includegraphics[scale = 0.6]{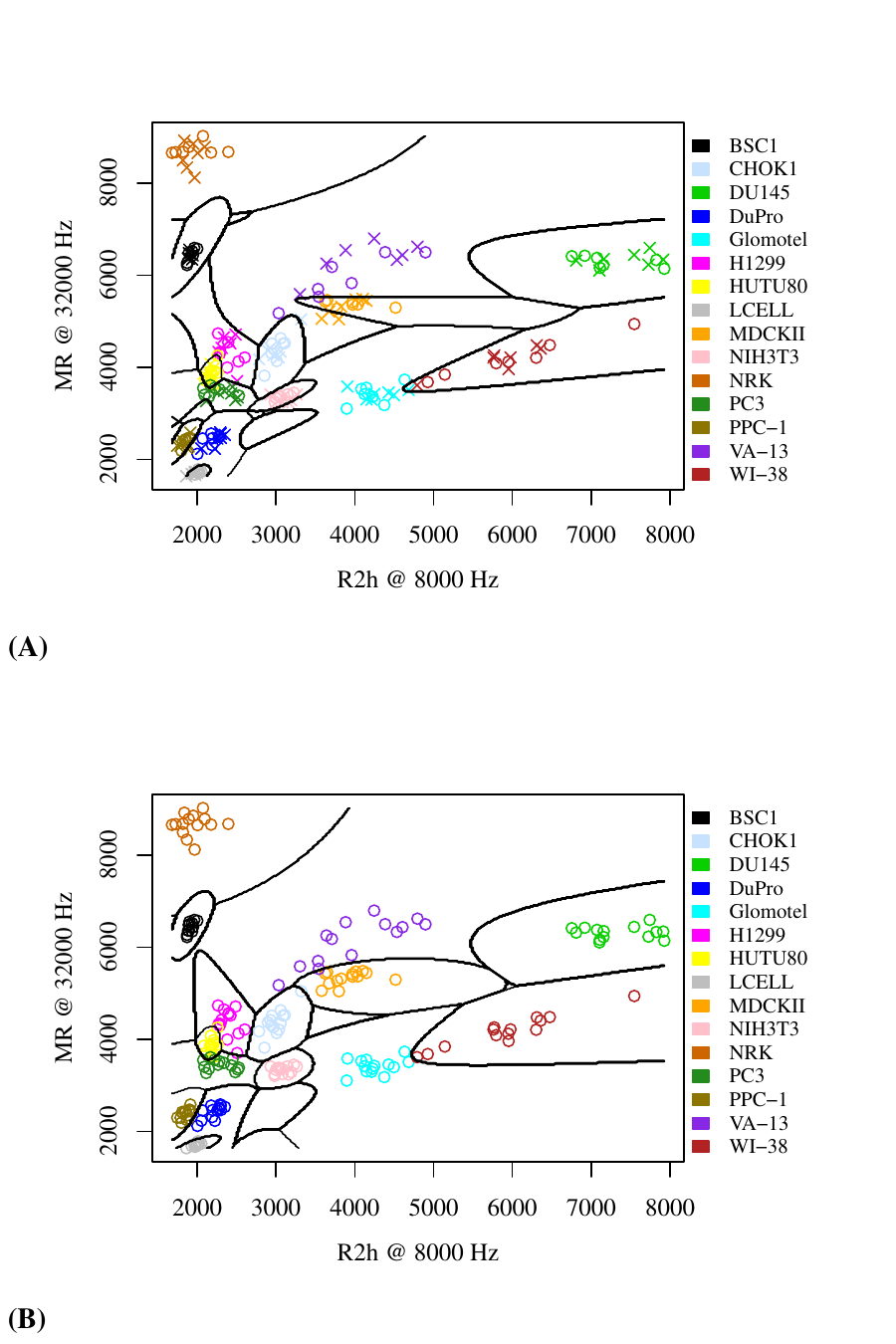}
\end{center}
\caption{{\bfseries QDA Visual Results:} Results obtained using best pair of features for cell line classification as determined through QDA with cells grown on gel serum.  (A) Example of training/testing set used to conduct classification analysis.  Training points are represented by hollow circles, test points represented by x's.  This separation corresponds to about 96\% predictive accuracy, as seen in Supplemental Table~\ref{tab:qda}.  (B) Boundaries created when using all 14 observations in training set (full information).}
\label{fig:qda_best}
\end{figure}

\newpage

\begin{figure}[h!]
\begin{center}
\includegraphics[width=\linewidth]{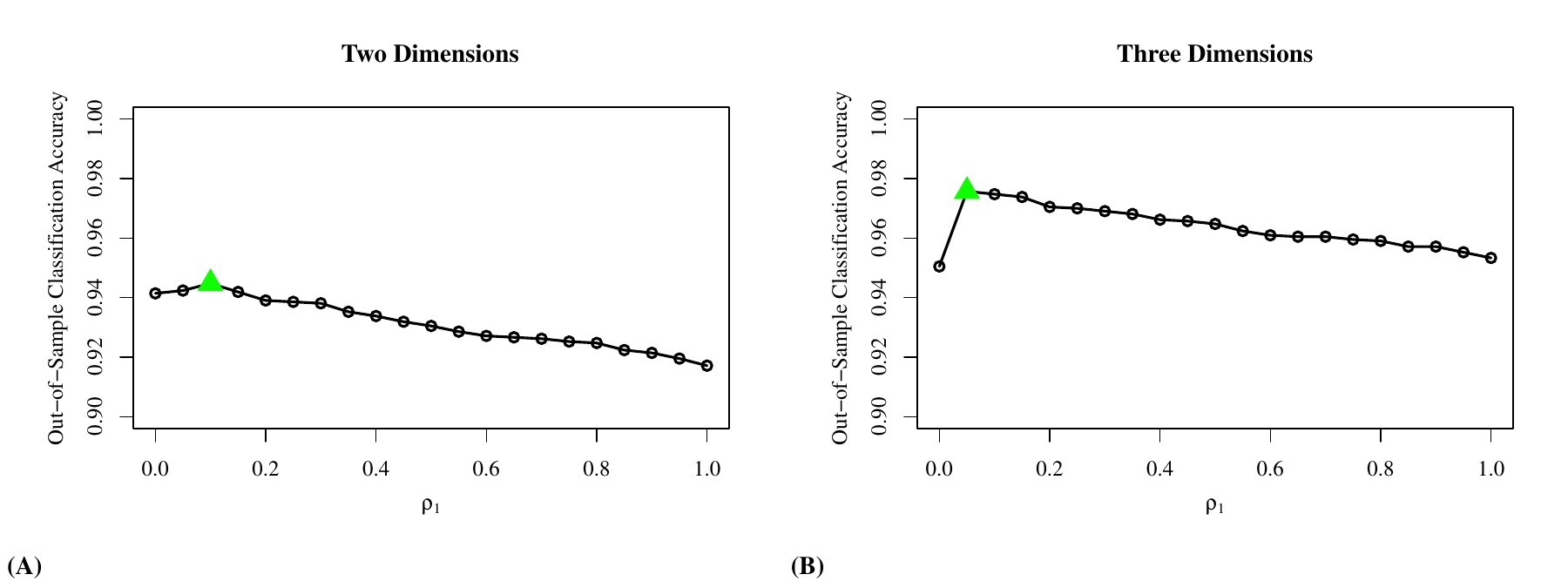}
\end{center}
\caption{{\bfseries Exploration of $\boldsymbol{\rho}_1$ for BSA-Based Cells:} Highest RDA out-of-sample classification accuracy for cells grown on BSA serum achieved among all  (A) pairs of features and (B) trio of features for each fixed value of $\rho_1 \in (0, 0.05, 0.1, ..., 0.95, 1)$, holding $\rho_2 = 0$. The maximum for each value of $\rho_1$ is associated with the best corresponding pair (trio) of features.  The maximum out-of-sample classification accuracy across all of the $\rho_1$ values is indicated with a green triangle in the figure. Note that $\rho_1 = 0$ corresponds to QDA and $\rho_1 = 1$ corresponds to LDA. The out-of-sample classification accuracy range in each panel is nearly the same, in magnitude, as the corresponding panel in Supplemental Figure~\ref{rho1s_gel}.  What has changed in this analysis are the values of $\rho_1$ associated with the maximum out-of-sample classification accuracy and the values of the these maxima --- they are uniformly lower than those seen in Supplemental Figure 1. These results suggest that RDA is slightly less effective at accurately classifying BSA based cells as opposed to gel based cells.}
\label{rho1s_bsa}
\end{figure}

\newpage

\begin{figure}[h!]
\begin{center}
\includegraphics[scale = .6]{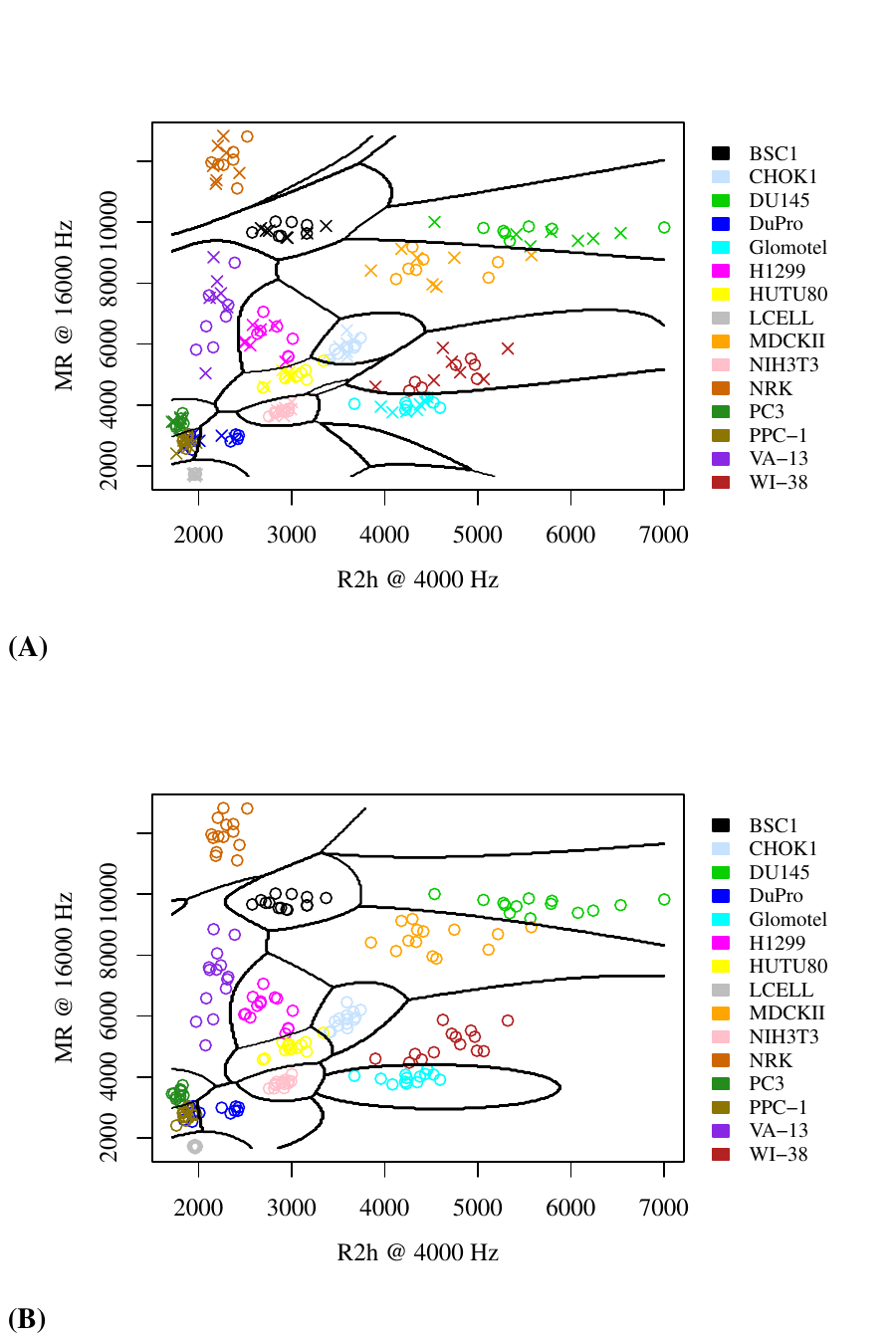}
\end{center}
\caption{{\bfseries RDA Visual Results for BSA-Based Cells:} Results obtained using best pair of features for cell line classification as determined through RDA with $\rho_1 = 0.1$ and $\rho_2 = 0$ with cells grown on BSA serum.  (A) Example of training/testing set used to conduct classification analysis. This separation corresponds to about 95\% predictive accuracy, as seen in Supplemental Table~\ref{tab:bsa}.  We see more overlap between cell lines here than we did for those cells inoculated in gel serum. (B)  Boundaries created when using all 14 observations in training set (full information).}
\label{fig:rda_best_bsa}
\end{figure}

\end{document}